\documentclass[prd,twocolumn, amsmath,amssymb,floatfix,nofootinbib]{revtex4}
\usepackage{mathrsfs,bm}
\usepackage{longtable,lscape}
\usepackage{txfonts}
\usepackage{indentfirst}
\usepackage{graphicx,color,dcolumn,booktabs}
\usepackage{multirow}
\usepackage{indentfirst}
\usepackage{multirow}
\usepackage{epsfig}
\usepackage{graphicx,color,dcolumn}
\usepackage{epstopdf}
\usepackage{amsmath}
\usepackage{diagbox}
\usepackage{changes}
\usepackage{threeparttable}
\usepackage{booktabs}
\usepackage{color}
\usepackage{amssymb}
\usepackage{cancel}
\definecolor{cover}{rgb}{0.77,0.87,0.88}
\definecolor{blueone}{rgb}{0.1,0.1,.7}
\definecolor{citec}{rgb}{0.14,0.47,0.09}
\definecolor{two}{rgb}{0.0,0.5,0.}
\definecolor{three}{rgb}{.5,.1,0.15}
\usepackage[bookmarks=true,bookmarksopen=false,plainpages=false,breaklinks=true,
   bookmarksnumbered=true,hypertexnames=false,
   filecolor=blue,urlcolor=three,menucolor=three,
   linkcolor=three,citecolor=blueone, colorlinks,
   anchorcolor=blue,runcolor=pink,frenchlinks=red
   pdfstartview=FitH,pdftitle=title,%
   pdfauthor=author]{hyperref}

\usepackage{relsize}
\usepackage{xspace}
\def\babar{\mbox{\slshape B\kern-0.1em{\smaller A}\kern-0.1em
    B\kern-0.1em{\smaller A\kern-0.2em R}}}

\newcolumntype{C}{>{$}c<{$}}
\AtBeginDocument{
\heavyrulewidth=.08em
\lightrulewidth=.05em
\cmidrulewidth=.03em
\belowrulesep=.65ex
\belowbottomsep=0pt
\aboverulesep=.4ex
\abovetopsep=0pt
\cmidrulesep=\doublerulesep
\cmidrulekern=.5em
\defaultaddspace=.5em
}

\allowdisplaybreaks[4]

\begin{document}
\title{Production of open-charm pentaquark molecules in decay $B^0 \rightarrow \bar{D}^0 p \bar{p}$}
\author{Shu-Yi Kong$^{1,2}$, Jun-Tao Zhu$^2$, Shu Chen$^1$, Jun He$^{1}$\footnote{Corresponding author: junhe@njnu.edu.cn}}

\affiliation{$^1$School of Physics and Technology, Nanjing Normal University, Nanjing 210097, China\\
$^2$School of Microelectronics and Control Engineering, Changzhou University, Changzhou 213164, China}

\date{\today}
\begin{abstract}

This study investigates the production of open-charm pentaquark molecular states, specifically $N\bar{D}^*$ and $\bar{N}\bar{D}^*$, within the $B^0 \rightarrow \bar{D}^0 p \bar{p}$ decay. We examine the invariant mass spectra of $p\bar{D}^0$ and $\bar{p}\bar{D}^0$, incorporating the rescattering process through a quasipotential Bethe-Salpeter equation approach. Our findings suggest the possible identification of the isoscalar $\bar{N}\bar{D}^*$ molecule with $3/2^+$ as the antiparticle partner of $\Lambda_c(2940)$ in the $\bar{p}\bar{D}^0$ mass spectrum. Although a molecular state near the $\bar{N}\bar{D}$ threshold with $1/2^-$ exists, its signal is weak, indicating that the $B^0 \rightarrow \bar{D}^0 p \bar{p}$ decay may not be ideal for its detection. A distinct signal of the isovector $N\bar{D}^*$ molecule with $1/2^-$ may appear in the $p\bar{D}^0$ invariant mass spectrum, while the signal for the $3/2^-$ state remains very weak. We emphasize the importance of the three-body decay of the bottom meson as a valuable method for studying open-charm molecules and advocate for increased attention and more precise experimental measurements of the $B^0 \rightarrow \bar{D}^0 p \bar{p}$ decay.

\end{abstract}

\maketitle

\section{INTRODUCTION}

Since the discovery of the $X(3872)$ by the Belle collaboration in
2003~\cite{Belle:2003nnu}, a series of new hadron states near threshold have
been reported by various experiments. Many of these states challenge
conventional classification as either three-quark baryons or quark-antiquark
mesons within the conventional quark model. Given their closeness to threshold
energies, a widely accepted hypothesis to explain these exotic hadrons is the
molecular state picture, where they are considered loosely bound states of two
hadrons. Beyond the $XYZ$ particles, hidden-charm pentaquarks, both with and
without strangeness, offer a rich spectrum of molecular states composed of a
charmed meson and a charmed baryon. In the open-charm sector, for systems
involving a nucleon and a charmed meson, several candidates for molecular states
have also been observed.

The series of experiments began with the discovery of the isotriplet
$\Sigma_c(2800)^{0,+,++}$ in 2005~\cite{Belle:2004zjl}. Initially identified in
the $\Lambda_c\pi^{-,0,+}$ mass spectrum by the Belle Collaboration, the
charge-neutral $\Sigma_c(2800)^{0}$ was later confirmed by the Babar
Collaboration~\cite{BaBar:2008get}. The spin parity $J^P$ of $\Sigma_c(2800)$
remains undetermined, and there is a notable discrepancy in mass measurements
between the two collaborations. However, both measurements place the mass close
to the $ND$ threshold. Another relevant structure near the $ND^*$ threshold is
the $\Lambda_{c}(2940)$, reported by the Babar Collaboration in
2007~\cite{BaBar:2006itc}. This state was observed in the $pD^{*0}$ invariant
mass spectrum with a mass of $m = 2939.8 \pm 1.3 \pm 1.0$~MeV and a width of
$\Gamma = 17.5 \pm 5.2 \pm 5.9$~MeV. The Belle Collaboration subsequently
reported their observation of this state in the $\Sigma^{0,++}_c\pi^{\pm}$
invariant mass spectrum, measuring a mass of $m = 2938.0 \pm 1.3
^{+2.0}_{-4.0}$~MeV and a width of $\Gamma = 17.5
^{+8}_{-5}{}^{+27}_{-7}$~MeV\cite{Belle:2006xni}.  In 2017, the LHCb
Collaboration determined the spin parity of $\Lambda_c(2940)$ as $3/2^-$ through
an amplitude analysis of the $\Lambda_b \rightarrow D^0 p \pi^-$
process~\cite{LHCb:2017jym}. The measured mass at LHCb, $m = 2944.8 \pm
1.3^{+3.5}_{-2.5}$~MeV, is consistent with the values reported by other
observations. Recently, the Belle Collaboration also studied the process
$\bar{B}^0 \rightarrow \Sigma_c^{0,++}\pi^{\pm}\bar{p}$ and identified a new
near-threshold structure, $\Lambda_c(2910)$, in the $\Sigma^{0,++}_c\pi^{\pm}$
mass spectrum, with a mass of $m = 2913.8 \pm 5.6 \pm 3.8$~MeV and a decay width
of $\Gamma = 51.8 \pm 20.0 \pm 18.8$MeV~\cite{Belle:2022hnm}.

The closeness of the masses of $\Sigma_c(2800)$, $\Lambda_c(2940)$, and
$\Lambda_c(2910)$ to the $N\bar{D}^{(*)}$ threshold has naturally led to
molecular explanations for understanding their internal structures. Although
debates persist regarding their spin parity, $\Sigma_c(2800)$ and
$\Lambda_c(2940)$ are frequently interpreted as molecular states of $ND$ and
$ND^*$, respectively~\cite{Wang:2020dhf,Liu:2023huu}. Several studies have
classified $\Sigma_c(2800)$ as a $ND$ molecule with spin parity $J^P =
1/2^-$~\cite{Jimenez-Tejero:2009cyn,Zhang:2012jk,Wang:2018jaj}. However, other
analyses estimate the strong two-body decay widths of decay $\Sigma_c \rightarrow
\Lambda_c\pi$ and suggest alternative spin parity of $1/2^+$ or
$3/2^-$~\cite{Dong:2009tg}.  For $\Lambda_c(2940)$, potential spin-parity
assignments within the molecular picture include $1/2^+$, $1/2^-$, or
$3/2^-$~\cite{Zhang:2012jk,Dong:2009tg,He:2006is,Dong:2010xv,He:2010zq,Ortega:2012cx,Entem:2016lzh,Garcia-Recio:2008rjt}.
A systematic study of the interaction between $D^*$ and the nucleon using a
one-boson-exchange model suggests that $\Lambda_c(2940)$ could be an isoscalar
$ND^*$ molecule with $1/2^+$ or $3/2^-$~\cite{He:2010zq}. Although limited
discussion on $\Lambda_c(2910)$ exists in the literature, this state has been
interpreted as an isoscalar open-charm molecular state of $ND^*$ with $J^P =
3/2^-$, based on QCD sum rules~\cite{Xin:2023gkf,Ozdem:2023eyz}.  Additionally,
the $\Lambda_c(2595)$ and $\Lambda_c(2625)$ have also been associated with $ND$
and $ND^*$ bound
states~\cite{Hofmann:2005sw,Garcia-Recio:2008rjt,Liang:2014kra,Lu:2014ina}. For
instance, Ref.~\cite{Garcia-Recio:2008rjt} classifies $\Lambda_c(2595)$ as a
quasi-bound $ND^*$ state, while Ref.~\cite{Liang:2014kra} shows that
$\Lambda_c(2595)$ couples to both $ND$ and $ND^*$, and $\Lambda_c(2625)$ couples
to $ND^*$. However, Ref.~\cite{Lu:2014ina} argues that $\Lambda_c(2595)$ is
predominantly a $\Sigma_c\pi$ state, with $\Lambda_c(2625)$ primarily coupling
to $\Sigma_c^*\pi$.

As of now, the connection between these resonances and the $ND^{(*)}$ molecular states remains uncertain. Further experimental insights from alternative production channels will be crucial for understanding the nature of these molecular states. The open-charm pentaquark, which contains a single $c$ or $\bar{c}$ quark, can be produced through the weak transition of the initial bottom hadron's $b$ ($\bar{b}$) quark to the $c$ ($\bar{c}$) quark via $W$ emission.
As previously mentioned, several candidates for these molecular states have been identified in the decay of bottom hadrons. Here, we propose that the three-body decay of bottom mesons serves as a promising platform for investigating open-charm molecules. Notably, the process $B^0 \rightarrow \bar{D}^0 p \bar{p}$ has a significant decay branching fraction of $\mathcal{B}_{B^0 \rightarrow \bar{D}^0 p \bar{p}} = (1.04 \pm 0.07) \times 10^{-4}$, according to the Review of Particle Physics (PDG)~\cite{Workman:2022ynf}. This substantial branching fraction highlights the importance of examining the invariant mass distribution of $\bar{p}\bar{D}^0$ to verify the potential existence of the open-charm molecule $\bar{N}\bar{D}^*$, the antiparticle partner of the $\Lambda_c(2940)$.
Furthermore, within the same three-body decay, the invariant mass distribution of ${p}\bar{D}^0$ allows us to explore the presence of the open-charm molecule $N\bar{D}^*$. Simultaneous observations of the $\bar{N}\bar{D}^*$ and $N\bar{D}^*$ molecular states could enhance our understanding of such states. Additionally, we will discuss the possible effects of the $\bar{N} \bar{D}$ and $N \bar{D}$ interactions in the energy region near the thresholds.

The structure of the paper is outlined as follows: after the introduction, we present the mechanism of the three-body decay. The potential kernels are constructed using heavy quark and chiral symmetries, along with effective Lagrangians under SU(3) symmetries. Additionally, a brief overview of the quasipotential Bethe-Salpeter equation (qBSE) approach is included. In Section~\ref{Sec:results}, we present explicit numerical results. Finally, a concise summary is provided in the last section.

\section{Theoretical frame}\label{Sec: Formalism}

\subsection{Rescattering mechanism of $B^0 \rightarrow \bar{D}^0 p \bar{p} $ three-body decay}

\begin{figure}[h!]
  \centering
  \includegraphics[scale=0.57,bb=86 608 525 750,clip]{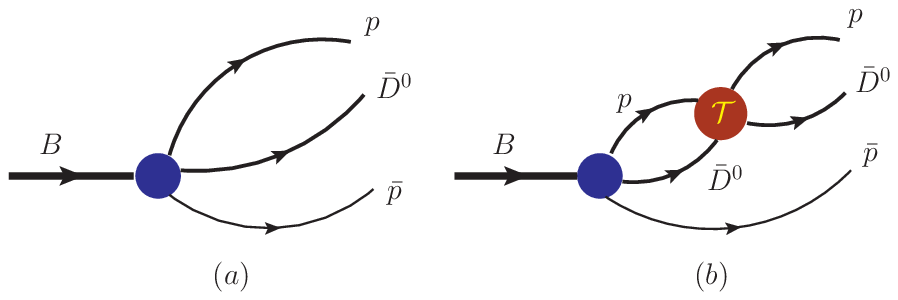}
  \caption{
  The diagram illustrates the process of $B^0 \rightarrow \bar{D}^0 p \bar{p}$
  for direct three-body decay (a), and the rescattering in the $p\bar{D}^0$
  channel (b). }
  \label{Fig: diagram}
\end{figure}

In the three-body decay process $B^0 \rightarrow \bar{D}^0 p \bar{p}$, the initial $B^0$ meson decays into $\bar{D}^{0} p \bar{p}$. Following this, rescattering takes place between two of the three final particles, leading to structures in the corresponding invariant mass spectrum. When the interaction is sufficiently strong and attractive, molecular states may form. This study focuses on rescattering in the $p\bar{D}^0$ and $\bar{p}\bar{D}^0$ channels. The rescattering process, particularly in the ${p}\bar{D}^0$ channel, is illustrated in Fig.~\ref{Fig: diagram}, while the $\bar{p}\bar{D}^0$ channel behaves analogously. Additionally, the coupling between $p\bar{D}^0$ and $p\bar{D}^{*0}$ is considered during the rescattering process.

The differential decay width of the initial $B^0$ meson is given by
\begin{align}
d\Gamma&=\frac{(2\pi)^4}{2m_{B^0}}|\mathcal{M}|^2d{\Phi}, \nonumber\\
d\Phi&=\delta^4(P-\sum_{i=1}^3p_i)\prod_{i=1}^{3}\frac{d^3{p}_i}{(2\pi)^32E_i},
\end{align}
where $\mathcal{M}$ represents the decay amplitude, $m_{B^0}$ is the mass of the $B^0$ meson, $P$ is the momentum of the initial $B^0$ meson, and $p_i$ and $E_i$ are the momentum and energy of the final particles.

We need to rewrite the Lorentz-invariant phase space in the center-of-mass frame of particles 1 and 2 as
\begin{align}
 d\Phi&=\frac{1}{8{(2\pi)}^9 m_{B^0}}|\boldsymbol{p}_1^{cm}||\boldsymbol{p}_3|d\Omega_1^{cm}d\Omega_3 dm_{12},
\end{align}
where $(|\boldsymbol{p}_1^{cm}|,\Omega_1^{cm})$ is the momentum of particle 1 in the rest frame of particles 1 and 2, and $(|\boldsymbol{p}_3|,\Omega_3)$ is the momentum of particle 3 in the rest frame of the decaying $B^0$ meson. Here, $m_{12}=\sqrt{(p_1+p_2)^2}$.
The $|\boldsymbol{p}_1^{cm}|$ and $|\boldsymbol{p_3}|$ are given by
\begin{align}
|\boldsymbol{p}_1^{cm}|&=\frac{1}{2m_{12}}\lambda^{\frac{1}{2}}(m^2_{12},m^2_{1},m^2_{2}),\\
|\boldsymbol{p}_3|&=\frac{1}{2m_{B^0}}\lambda^{\frac{1}{2}}({m_{B^0}}^2,m^2_{12},m^2_{3}),
\end{align}
with  $\lambda(a,b,c)=a^2+b^2+c^2-2ab-2ac-2bc$, and $m_{1,2,3}$ are the masses of final particles.

Thus, the differential decay width of the $B^0$ meson in the center-of-mass frame is
\begin{align}
 d\Gamma &=\frac{1}{(2\pi)^5}\frac{1}{16m_{B^0}^2}|\mathcal{M}|^2 |\boldsymbol{p}_1^{cm}||\boldsymbol{p}_3|d\Omega^{cm}_1 d\Omega_3 dm_{12}.
\end{align}

As illustrated in Fig.\ref{Fig: diagram}, the decay process of the $B^0$ meson
involves two main processes: direct decay and rescattering. The Lagrangian for
$B^0 \rightarrow \bar{D}^0 p \bar{p}$ can be expressed as: \begin{align}
\mathcal{L}_{B^0 \rightarrow\bar{D} p \bar{p}} &= \frac{1}{2} g_A
B^0\bar{p}\gamma^{\mu} \partial_{\mu}{\bar{D}}p-\frac{1}{2} g_B
B^0\bar{p}\gamma^{\mu}\gamma^{5} \partial_{\mu}{\bar{D}}p, \end{align} where
$g_A$ and $g_B$ represent the parity-violating and parity-conserving coupling
constants, respectively. To simplify the calculations, we assume $g_A = g_B =
g_1$. Although this introduces some uncertainties due to the unknown exact
values of these constants, the subsequent renormalization of the results mean
that the choice of $g_1$ serves primarily as an indicative value. The
sensitivity of this assumption will be tested in Section \ref{Sec:results} by setting
$g_A = 0$ or $g_B = 0$.

Now, the task is to determine the coupling constant $g_1$. Experimentally, the decay width of this three-body decay channel can be inferred from the decay branching fraction $\mathcal{B}_{B^0 \to \bar{D}^0 p \bar{p}} = (1.04 \pm 0.07) \times 10^{-4}$ and the decay width of $B^0$ derived from its lifetime, $\Gamma_{B^0} = 4.325 \times 10^{-10}$ MeV~\cite{Workman:2022ynf}. The differential width of the initial $B^0$ meson can also be written as,
\begin{eqnarray}
d\Gamma &=&\frac{1}{256{\pi}^2}\frac{|\mathcal{M}|^2}{m_{B^0}^3}{dm^2_{12}dm^2_{23}},~\label{dwidth}
\end{eqnarray}
where $|\mathcal{M}|^2$ is the modulus squared of the $B^0$ meson direct decay amplitude, and $m_{23}=\sqrt{(p_2+p_3)^2}$.

The decay amplitude $\mathcal{M}$ for the process $B^0 \rightarrow \bar{D}^0 p \bar{p}$ can be written using the Lagrangian as,
\begin{eqnarray}
\mathcal{M}_{B^0 \rightarrow\bar{D}^0 p \bar{p}}&=& \frac{g_1}{2} \bar{v}_3\not{k}_{2}
(1-\gamma^{5}) u_1,
\end{eqnarray}
where $v_3$ and $\bar{u}_1$ represent the spinors for the antiproton ($\bar{p}$) and proton ($p$) as particles 3 and 1, respectively.
The ${k}_{2}$ refers to the momentum of the $\bar{D}$ meson as particle 2. The total differential decay width of the initial $B^0$ meson can then be expressed as,
\begin{eqnarray}
\Gamma&=&\frac{1}{256\pi^3 m_{B^0}^3} \int_{(m_1+m_2)^2}^{(m_{B^0}-m_3)^2} dm^2_{12}
 \int^{m^2_{23 max}}_{m^2_{23 min}} |\mathcal{M}|^2 dm^2_{23}.~\label{intdgamma}
\end{eqnarray}

According to the constraints mentioned earlier, we can integrate the differential decay width, leading to the result $\Gamma_{B^0\to {\bar{D}^0 p \bar{p}}} = 0.016\times g_1^2$ MeV. Meanwhile, the decay width $\Gamma_{B^0\to {\bar{D}^0 p \bar{p}}}$ can be determined experimentally from the decay branching ratio $\mathcal{B}_{{B^0}\to {\bar{D}^0 p \bar{p}}}$ and the lifetime of the $B^0$ meson as follows,
\begin{eqnarray}
\mathcal{B}_{{B^0}\to {\bar{D}^0 p \bar{p}}} &=&\frac{\Gamma_{B^0\to {\bar{D}^0 p \bar{p}}}}
{\Gamma_{B^0}}= (1.04 \pm 0.07 ) \times 10^{-4}, \nonumber\\
\Gamma_{B^0\to {\bar{D}^0 p \bar{p}}}&=&4.498\times 10^{-14}~\mathrm{MeV}.
\end{eqnarray}

From this, we can deduce the coupling constant $g_1^2 = 2.81 \times 10^{-12}$.
It is important to
note that the decay width for $B^0 \rightarrow \bar{D}^0 p \bar{p}$ should
account for contributions from direct decay, rescattering, and other processes
not considered here, such as those via triangle diagrams~\cite{Guo:2019twa}. Therefore, the value of $g_1$ obtained
should be viewed as an upper bound, serving as a reference for calculation.
Further discussion will follow once the results are obtained.

\subsection{Lagrangians and potential kernel}

The potential kernel  ${\cal V}$ of the rescattering effect can be constructed by using heavy quark and chiral symmetries, along with SU(3) symmetries. The couplings of light mesons to anticharmed mesons can be described using the following effective Lagrangians~\cite{Cheng:1992xi,Yan:1992gz,Wise:1992hn,Burdman:1992gh,Casalbuoni:1996pg},
\begin{eqnarray}
\mathcal{L}_{\widetilde{\mathcal{P}}^*\widetilde{\mathcal{P}}^*\mathbb{P}} &=&i \frac{g}{f_\pi m_{\widetilde{\mathcal{P}}^*}}\varepsilon_{\alpha\mu\nu\lambda}
i\overleftrightarrow{\partial}^\alpha\widetilde{\mathcal{P}}^{*\mu\dag}_{a}\widetilde{\mathcal{P}}^{*\lambda}_{b}
\partial^\nu{}\mathbb{P}_{ab},\nonumber\\
\mathcal{L}_{\widetilde{\mathcal{P}}^*\widetilde{\mathcal{P}}\mathbb{P}} &=&\frac{2g}{f_\pi}(\widetilde{\mathcal{P}}^{*\dag}_{a\lambda}\widetilde{\mathcal{P}}_b+
\widetilde{\mathcal{P}}^{\dag}_{a}\widetilde{\mathcal{P}}^{*}_{b\lambda})\partial^\lambda{}\mathbb{P}_{ab},\nonumber\\
\mathcal{L}_{\widetilde{\mathcal{P}}\widetilde{\mathcal{P}}\mathbb{V}}
  &=&\frac{\sqrt{2}\beta g_V}{2m_{\widetilde{\mathcal{P}}}}\widetilde{\mathcal{P}}^{\dag}_a
  \widetilde{\mathcal{P}}^{}_b
i\overleftrightarrow{\partial}\cdot\mathbb{V}_{ab},\nonumber\\
\mathcal{L}_{\widetilde{\mathcal{P}}^*\widetilde{\mathcal{P}}\mathbb{V}}
  &=&- \frac{\sqrt{2}\lambda{}g_V}{\sqrt{M_{\widetilde{\mathcal{P}}^*}m_{\widetilde{\mathcal{P}}}}}
i\overleftrightarrow{\partial}^\lambda\varepsilon_{\lambda\mu\alpha\beta}
(\widetilde{\mathcal{P}}^{*\mu\dag}_a\widetilde{\mathcal{P}}^{}_b
+
\widetilde{\mathcal{P}}^{\dag}_a\widetilde{\mathcal{P}}_b^{*\mu})
  (\partial^\alpha{}\mathbb{V}^\beta)_{ab},\nonumber\\
\mathcal{L}_{\widetilde{\mathcal{P}}^*\widetilde{\mathcal{P}}^*\mathbb{V}}
  &= &-\frac{\sqrt{2}\beta g_V}{2m_{\widetilde{\mathcal{P}}^*}}
  \widetilde{\mathcal{P}}^{*\dag}_a\cdot\widetilde{\mathcal{P}}_b^{*}
 i\overleftrightarrow{\partial}\cdot\mathbb{V}_{ab}\nonumber\\
  &-&i2\sqrt{2}\lambda{}g_V\widetilde{\mathcal{P}}^{*\mu\dag}_a\widetilde{\mathcal{P}}^{*\nu}_b(\partial_\mu{}
  \mathbb{V}_\nu - \partial_\nu{}\mathbb{V}_\mu)_{ab},\nonumber\\
\mathcal{L}_{\widetilde{\mathcal{P}}\widetilde{\mathcal{P}}\sigma}
  &=&
 -2g_s\widetilde{\mathcal{P}}^{}_b\widetilde{\mathcal{P}}^{\dag}_b\sigma,\nonumber\\
  \mathcal{L}_{\widetilde{\mathcal{P}}^*\widetilde{\mathcal{P}}^*\sigma}
  &= &2g_s\widetilde{\mathcal{P}}^{*}_b\cdot{}\widetilde{\mathcal{P}}^{*\dag}_b\sigma,\label{Eq:L}
\end{eqnarray}
where the anticharmed meson $\widetilde{\mathcal{P}}^{(*)}=(\bar{D}^{(*)0},D^{(*)-},D^{(*)-}_s)$ satisfy the normalization relations $\langle
0|{\widetilde{\mathcal{P}}}|\bar{Q}q(0^-)\rangle
=\sqrt{M_\mathcal{P}}$ and $\langle
0|\widetilde{\mathcal{P}}^*_\mu|\bar{Q}{q}(1^-)\rangle=
\epsilon_\mu\sqrt{M_{\mathcal{P}^*}}$.
The matrices $\mathbb P$ and $\mathbb V$ represent the pseudoscalar ($P$) and vector ($V$) mesons, respectively, and are given by
\begin{eqnarray}
    {\mathbb P}=\left(\begin{array}{ccc}
        \frac{\sqrt{3}\pi^0+\eta}{\sqrt{6}}&\pi^+&K^+\\
        \pi^-&\frac{-\sqrt{3}\pi^0+\eta}{\sqrt{6}}&K^0\\
        K^-&\bar{K}^0&-\frac{2\eta}{\sqrt{6}}
\end{array}\right),
\mathbb{V}=\left(\begin{array}{ccc}
\frac{\rho^0+\omega}{\sqrt{2}}&\rho^{+}&K^{*+}\\
\rho^{-}&\frac{-\rho^{0}+\omega}{\sqrt{2}}&K^{*0}\\
K^{*-}&\bar{K}^{*0}&\phi
\end{array}\right).\label{MPV}
\end{eqnarray}

The parameters mentioned above are listed in Table~\ref{coupling constant1}, which are cited from the literature~\cite{Falk:1992cx,Isola:2003fh,Liu:2009qhy,Chen:2019asm}.
\renewcommand\tabcolsep{0.45cm}
\renewcommand{\arraystretch}{1.5}
\begin{table}[h!]
\caption{Coupling constants in the heavy quark and chiral symmetries theory. The $\lambda$ and $f_{\pi}$ are in units of GeV$^{-1}$. Other parameters are in units of 1.} \label{coupling constant1}
\begin{tabular}{cccccc}\bottomrule[1.5pt]
$\beta$&$g_s$&$g_V$&$f_{\pi}$&$g$&$\lambda$\\\hline
0.9 &0.76&3.25&0.132&0.9&0.56\\
\bottomrule[1.5pt]
\end{tabular}
\end{table}

The explicit effective Lagrangians for the couplings of nucleons with light mesons are given by~\cite{Ronchen:2012eg,Kamano:2008gr,Zhao:2013ffn},
\begin{eqnarray}
\mathcal{L}_{NNP}&=& -\frac{g_{NNP}}{m_P}\bar{N}\gamma^{5}\gamma^{\mu}\partial_{\mu}PN,\nonumber\\
\mathcal{L}_{NNV}&=&-\bar{N}\left[g_{NNV}\gamma^{\mu}-\frac{f_{NNV}}{2m_{N}}\sigma^{\mu\nu}\partial_{\nu}\right]V_{\mu}N,\nonumber\\
\mathcal{L}_{NN\sigma}&=&-g_{NN\sigma}\bar{N}\sigma N. \label{Eq: NNm}
\end{eqnarray}
The $m_P$ and $m_V$ are the masses of the pseudoscalar and vector mesons, respectively, as those defined in Eq.~(\ref{MPV}), while $m_N$ represents the nucleon mass. The values of the masses considered in the current work are provided in Table~\ref{masses}.
\renewcommand\tabcolsep{0.37cm}
\renewcommand{\arraystretch}{1.5}
\begin{table}[h!]
\caption{The masses of the light mesons considered in the current work and nucleon. The units are all GeV.
\label{masses}}
\begin{tabular}{cccccc}\bottomrule[1.5pt]
$\pi$&$\eta$&$\rho$&$\omega$&$\sigma$ & $N$\\\hline
0.137 &0.547&0.775&0.782&0.550 & 0.938\\
\bottomrule[1.5pt]
\end{tabular}
\end{table}

The parameters are listed in Table~\ref{coupling constant}, and they adhere to the SU(3) symmetries~\cite{Ronchen:2012eg,Zhao:2013ffn,deSwart:1963pdg,Lu:2020qme,Zhu:2022fyb,Kong:2022rvd,Kong:2023dwz}.
\renewcommand\tabcolsep{0.27cm}
\renewcommand{\arraystretch}{1.5}
\begin{table}[h!]
\caption{Coupling constants in effective Lagrangians in Eq.~(\ref{Eq: NNm}). All constants are in units of 1. \label{coupling constant}}
\begin{tabular}{ccccccccccccccccccc}\bottomrule[1.5pt]
$g_{NN\pi}$&$g_{NN\eta}$&$g_{NN\rho}$&$f_{NN\rho}$&$g_{NN\omega}$&$f_{NN\omega}$&$g_{NN\sigma}$\\\hline
0.989 &0.346&3.25&19.825&11.7&0&6.59\\
\bottomrule[1.5pt]
\end{tabular}
\end{table}

With the given Lagrangians for the vertices, the potential kernel for the rescattering process can be constructed by using the one-boson-exchange model, following standard Feynman rules, as outlined in Refs.~\cite{He:2019ify,He:2015mja},
\begin{eqnarray}%
{\cal V}_{{P},\sigma}=I_{{P},\sigma}\Gamma_1\Gamma_2 P_{{P},\sigma}(q^2),\quad
{\cal V}_{{V}}=I_{{V}}\Gamma_{1\mu}\Gamma_{2\nu}  P^{\mu\nu}_{{V}}(q^2),~\label{fmv}
\end{eqnarray}
where $\Gamma_{1(\mu)}$ and $\Gamma_{2(\nu)}$ represent the upper and lower vertices of the interaction via meson exchanges. The propagators of the exchanged mesons are defined as follows,
\begin{align}%
P_{P,\sigma}(q^2)&=
\frac{i}{q^2-m_{P,\sigma}^2}~f_i(q^2),\nonumber\\
P^{\mu\nu}_{V}(q^2)&=i\frac{-g^{\mu\nu}+q^\mu
q^\nu/m^2_{{V}}}{q^2-m_{V}^2}~f_i(q^2),
\end{align}
where the form factor $f_i(q^2)$ is introduced to account for the off-shell
effects of the exchanged meson, typically expressed as $e^{-(m_e^2 - q^2)^2 /
\Lambda_e^4}$  with $m_e$, $q$, and $\Lambda_e$ representing the mass, momentum,
and cutoff parameter of the exchanged light mesons ($e$), respectively. The exchanged
mesons include pseudoscalar ($P=\pi$ and $\eta$ ), vector ($V=\rho$ and
$\omega$), and scalar ($\sigma$) mesons.

The $I_{{P},{V},\sigma}$ are the flavor factors corresponding to the specific meson exchanges. The flavor factors for $N\bar{D}^{(*)}$ interactions are listed in Table~\ref{flavor factor}. If the nucleon $N$ is replaced with an antinucleon $\bar{N}$, the flavor factors for $\bar{N}\bar{D}^{(*)}$ interactions can be derived using the well-known G-parity rule~\cite{PHILLIPS:1967wls,Klempt:2002ap}.
\renewcommand\tabcolsep{0.5cm}
\renewcommand{\arraystretch}{1.6}
\begin{table}[h!]
 \centering
\caption{Flavor factors $I_e$ for $N\bar{D}^{(*)}$ interactions. The value of $I_\sigma$ should be 0 for couplings between different channels, and vertices involving three pseudoscalar mesons should be forbidden.\label{flavor factor}}
\begin{tabular}{c|ccccccc}\toprule[1.5pt]
$I$& $\pi$& $\eta$ &$\rho$&$\omega$  &$\sigma$ \\\hline
$0$ &$-\frac{3\sqrt{2}}{2}$&$\frac{\sqrt{6}}{6}$&$-\frac{3\sqrt{2}}{2}$ &$\frac{\sqrt{2}}{2}$ & $1$\\
$1$&$\frac{\sqrt{2}}{2}$&$\frac{\sqrt{6}}{6}$&$\frac{\sqrt{2}}{2}$ &$\frac{\sqrt{2}}{2}$ & $1$\\
\bottomrule[1.5pt]
\end{tabular}
\end{table}

With the above information, the explicit expressions for the potentials can be constructed as follows,
\begin{align}
i{\cal V}_{{V}}^{N\bar{D}\to N\bar{D}}&=-I_{{V}} \frac{\sqrt{2}}{2}\beta g_V {(k_{2i}+k_{2f})}_\mu
\frac{g^{\mu \nu}+q^\mu q^\nu/m_V^2}{q^2-m_V^2} \nonumber\\
&\cdot \bar{u}\left[-g_{NNV}\gamma_\nu+\frac{f_{NNV}}{4m_N}(\gamma_\nu \rlap{$\slash$} q-\rlap{$\slash$} q \gamma_\nu)\right]u f_i(q^2),\nonumber\\
%
i{\cal V}_{\sigma}^{N\bar{D}\to N\bar{D}}&=2 I_{{V}} g_s g_{NN\sigma} m_{\bar{D}} \frac{1}{{q^2}-m_{\sigma}^2} \bar{u} u f_i(q^2),\nonumber\\
i{\cal V}_{{P}}^{N\bar{D}^*\to N\bar{D}^*}&=\frac{-i I_{{P}}g g_{NNP}}{f_{\pi} m_{P}}\varepsilon_{\lambda\mu\alpha\beta}{(k_{2i}+k_{2f})}^\alpha \epsilon^{\dag\mu} \epsilon^\lambda q^\nu \nonumber\\
&\cdot\frac{1}{{q^2}-m_{P}^2}\bar{u} \gamma^5 \rlap{$\slash$} q u f_i(q^2),\nonumber\\
i{\cal V}_{{V}}^{N\bar{D}^*\to N\bar{D}^*}
&=I_{{V}}[\frac{\sqrt{2}}{2}\beta g_V~\epsilon^{\dag}\cdot\epsilon~{(k_{2f}+k_{2i})}_\mu+2\sqrt{2}\lambda\nonumber\\
&\cdot g_V m_{\bar{D}^*}(\epsilon^{\dag}\cdot q ~\epsilon_{\mu}-\epsilon^\dag_{\mu}~\epsilon\cdot q)]
\frac{-g^{\mu \nu}+q^\mu q^\nu/m_V^2}{q^2-m_V^2}\nonumber\\
&\cdot\bar{u}\left[g_{NNV}\gamma_\nu+\frac{f_{NNV}}{4m_N}(\gamma_\nu \rlap{$\slash$} q-\rlap{$\slash$} q \gamma_\nu)\right]u f_i(q^2),\nonumber\\
i{\cal V}_{\sigma}^{N\bar{D}^*\to N\bar{D}^*}
&=-2 I_{\sigma}g_s g_{NN\sigma} m_{\bar{D}^*} \epsilon\cdot \epsilon^{\dag} \frac{1}{{q^2}-m_{\sigma}^2} \bar{u} u f_i(q^2),\nonumber\\
i{\cal V}_{{P}}^{N\bar{D}^*\to N\bar{D}}
&=\frac{2 I_{{P}} g g_{NNP}}
{f_{\pi} m_{P}}\sqrt{m_{\bar{D}} m_{\bar{D}^*}}
~\epsilon\cdot q \frac{1}{{q^2}-m_{P}^2} \bar{u} u f_i(q^2),\nonumber\\
i{\cal V}_{{V}}^{N\bar{D}^*\to N\bar{D}}
&=-i I_{{V}}\sqrt2 \lambda g_V  \varepsilon_{\lambda\alpha\beta\mu}{(k_{2f}+k_{2i})}^\lambda
\epsilon^\alpha q^\beta \frac{-g^{\mu \nu}}{q^2-m_V^2}\nonumber\\
&\cdot\bar{u}\left[g_{NNV}\gamma_\nu+\frac{f_{NNV}}{4m_N}(\gamma_\nu \rlap{$\slash$} q-\rlap{$\slash$} q \gamma_\nu)\right]u f_i(q^2),\nonumber\\
i{\cal V}_{{P}}^{N\bar{D}\to N\bar{D}^*}&=\frac{2 I_{{P}} g g_{NNP}}
{f_{\pi} m_{P}}\sqrt{m_{\bar{D}} m_{\bar{D}^*}}
~\epsilon^{\dag}\cdot q\frac{1}{{q^2}-m_{P}^2} \bar{u} u f_i(q^2),\nonumber\\
i{\cal V}_{{V}}^{N\bar{D}\to N\bar{D}^*}
&=-i I_{{V}}\sqrt2 \lambda g_V  \varepsilon_{\lambda\alpha\beta\mu}{(k_{2f}+k_{2i})}^\lambda
\epsilon^{\dag\alpha} q^\beta \frac{-g^{\mu \nu}}{q^2-m_V^2}\nonumber\\
&\cdot\bar{u}\left[g_{NNV}\gamma_\nu+\frac{f_{NNV}}{4m_N}(\gamma_\nu \rlap{$\slash$} q-\rlap{$\slash$} q \gamma_\nu)\right]u f_i(q^2) ,
\end{align}
where $k_{2i}$ and $k_{2f}$ are the momenta for the initial and final particle 2, and the momenta for the exchanged meson here are defined as $q = k_{2f} - k_{2i}$. The $\epsilon$ and $\epsilon^\dag$ are the polarized vectors for the initial and final vector mesons, respectively. The $u$ and $\bar{u}$ represent the initial and final spinors of the nucleons.

\subsection{The qBSE approach}~\label{qbse}

The Bethe-Salpeter equation is a 4-dimensional relativistic kinematic equation
that can be used to treat two-body scattering. In our previous
works~\cite{He:2019ify,He:2015mja}, we employed a series of quasipotential
approximation methods to reduce the 4-dimensional Bethe-Salpeter equation to a
1-dimensional equation. The partial-wave rescattering amplitude ${\cal T}$ with
a certain spin parity $J^P$ can be expressed as,
\begin{align}
  &i{\cal T}^{J^P}_{\lambda'_1\lambda'_2,\lambda_1\lambda_2}({\rm p}',{\rm p})\nonumber\\
  &=i{\cal V}^{J^P}_{\lambda'_1\lambda'_2,\lambda_1\lambda_2}({\rm p}',{\rm
  p})+\sum_{\lambda''}\int\frac{{\rm
  p}''^2d{\rm p}''}{(2\pi)^3}\nonumber\\
  &\cdot
  i{\cal V}^{J^P}_{\lambda'_1\lambda'_2,\lambda''_1\lambda''_2}({\rm p}',{\rm p}'')
  G_0({\rm p}'')i{\cal T}^{J^P}_{\lambda''_1\lambda''_2,\lambda_1\lambda_2}({\rm p}'',{\rm
  p}),\quad\quad \label{Eq: BS_PWA}
\end{align}
where the sum extends only over independent helicities $\lambda$. $G_0({\rm p}'')$ is a reduced propagator with the spectator approximation in the center-of-mass frame as~\cite{He:2015mja},
\begin{align}
G_0&=\frac{1}{2E_h({\rm p''})[(W-E_h({\rm
p}''))^2-E_l^{2}({\rm p}'')]}.
\end{align}
In the spectator approximation, the heavier particle, denoted as $h$, is placed on-shell, with the zero component of four-momentum $p''_h$ given by $p''^0_h=E_{h}({\rm p}'')=\sqrt{m_{h}^{2}+\rm p''^2}$ with $m_h$ being the mass of the heavier particle. The corresponding zero component of four-momentum for the lighter particle, denoted as $l$, is then $W-E_{h}({\rm p}'')$, where $W$ represents the total energy of the system comprising particles 1 and 2. Here and hereafter, we define the value of the momentum as ${\rm p}=|{\bm p}|$.

The ${\cal V}_{\lambda'_1\lambda'_2,\lambda_1\lambda_2}^{J^P}$ can be obtained by decomposing the partial wave on the potential kernel constructed in Eq.~(\ref{fmv}), as detailed in~\cite{He:2015mja}.
\begin{align}
i{\cal V}_{\lambda'_1\lambda'_2,\lambda_1\lambda_2}^{J^P}({\rm p},{\rm p}')
&=2\pi\int d\cos\theta
~[d^{J}_{\lambda_{12}\lambda'_{12}}(\theta)
i{\cal V}_{\lambda'_1\lambda'_2,\lambda_1\lambda_2}({\bm p},{\bm p}')\nonumber\\
&+\eta d^{J}_{-\lambda_{12}\lambda'_{12}}(\theta)
i{\cal V}_{-\lambda'_1-\lambda'_2,\lambda_1\lambda_2}({\bm p},{\bm p}')],~\label{v1m}
\end{align}
where $\eta=PP_1P_2(-1)^{J-J_1-J_2}$, with $P$ and $J$ being parity and spin for particles 1 and 2. Here, $\lambda_{12}=\lambda_2-\lambda_1$, while ${\bm p}$ and ${\bm p'}$ denote the initial and final relative momenta, respectively. These are chosen as ${\bm p'}=(0,0,{\rm p'})$ and ${\bm p}=({\rm p}\sin\theta,0,{\rm p}\cos\theta)$ in the center-of-mass system of particles 1 and 2. The $d^J_{\lambda\lambda'}(\theta)$ represents the Wigner $d$-matrix. An exponential regularization is introduced as a form factor into the reduced propagator, given by $G_0({\rm p}'')\to G_0({\rm p}'')e^{-2(p''^2_l-m_l^2)^2/\Lambda_r^4}$, with the cutoff $\Lambda_r$ and the mass of the lighter constituent $m_l$ related as $\Lambda_r= m_l+\alpha_r 0.22$~GeV. The cutoff parameters $\Lambda_r$ and $\Lambda_e$ play analogous roles in the results. For simplification, we set $\Lambda_e = \Lambda_r$ in the current calculations, meaning a single parameter $\alpha$ will be used in the following discussion.

Then, the 1-dimensional integral equation~(\ref{Eq: BS_PWA}) can be transformed into a matrix equation by discretizing the momenta ${\rm p}'$, ${\rm p}$, and ${\rm p}''$ using Gauss quadrature, as~\cite{He:2015mja}.
\begin{align}
{T}_{ik}
={ V}_{ik}+\sum_{j=0}^N{ V}_{ij}G_j{ T}_{jk},\label{Eq: matrix}
\end{align}
where $i$, $k$, and $j$ are indices of momenta after discretization. We also integrate the helicities into such indices. The value of $N$ is the Gaussian discretization dimension, and its specific value depends on the stability of the calculation results. In the current work, the value of the discretization dimension is chosen as 10.

The propagator $G$ can be expressed as a $j$-dimensional diagonal matrix as
\begin{align}
	G_{j>0}&=\frac{w({\rm p}''_j){\rm p}''^2_j}{(2\pi)^3}G_0({\rm
	p}''_j), \nonumber\\
G_{j=0}&=-\frac{i{\rm p}''_o}{32\pi^2 W}+\sum_j
\left[\frac{w({\rm p}_j)}{(2\pi)^3}\frac{ {\rm p}''^2_o}
{2W{({\rm p}''^2_j-{\rm p}''^2_o)}}\right],
\end{align}
with the on-shell momentum ${\rm p}''_o=\lambda^{\frac{1}{2}}(W^2,m^2_1,m^2_2)/2W$. After continuing the energy $W$ into the complex plane, changing the sign of ${\rm p}''_o$ transforms the propagators from $G^{(I)}$ on the first Riemann sheet to $G^{(II)}$ on the second Riemann sheet, as described in Ref.~\cite{Roca:2005nm}.

\subsection{Invariant mass spectrum of molecular states}
After considering the direct decay and rescattering process above, the Lorentz-invariant amplitude of three-body decay of the $B^0$ meson with rescattering can be written in the center-of-mass frame of particles 1 and 2 as follows~\cite{He:2017lhy,Ding:2023yuo}:

\begin{align}
&{\cal M}_{\lambda_1,\lambda_2,\lambda_3;\lambda}(p_1,p_2,p_3)\nonumber\\
&=\sum_{\lambda_1,\lambda_2}\int \frac{d^4p'^{cm}_2}{(2\pi)^4}
{\cal T}_{\lambda_1\lambda_2}(p^{cm}_1,p^{cm}_2;p'^{cm}_1,p'^{cm}_2) \nonumber\\
&\cdot G(p'^{cm}_1,p'^{cm}_2){\cal A}_{\lambda_1\lambda_2,\lambda_3;\lambda}(p'^{cm}_1,p'^{cm}_2,p^{cm}_3,P^{cm}),
\end{align}
where the $p^{cm}_{1, 2, 3}$ and $P^{cm}$ are the momenta of the final particles and the $B^0$ in the center-of-mass frame of particles 1 and 2, and $\lambda_{1,2,3}$ and $\lambda$ are the helicities for the final and initial particles, respectively.

In section~\ref{qbse}, we conducted a partial wave expansion of the rescattering potential kernel ${\cal V}$. Similarly, we need to perform a partial wave expansion of the direct decay amplitude ${\cal A}$ as~\cite{Gross:2008ps},
\begin{align}
{\cal A}^{J}({\rm p}'^{cm}_2,\Omega^{cm}_2)&=N_J \int d\Omega^{cm}_2
 {\cal A}_{{\lambda}^{'}_{12}\lambda}
D^{J^*}_{\lambda,\lambda^{'}_{12}}( \Omega^{cm}_2),
\end{align}
where $N_J$ is a normalization constant equal to $\sqrt{(2J+1)/4\pi}$.
The spherical angle $\Omega_2^{cm}$ defines the orientation of the momentum of
particle 2 in the center-of-mass frame of particles 1 and 2. Hence, the
partial-wave amplitude is given by

\begin{align}
&{\cal M}^{J}_{\lambda_1,\lambda_2,\lambda_3;\lambda}(p_1,p_2,p_3)\nonumber\\
&=\sum_{J\lambda}N_{J}D^{J*}_{\lambda;\lambda_{12}}
( \Omega_2^{cm})\sum_{\lambda'_1\lambda'_2}\int \frac{{\rm p}'^{*2}_2d{\rm p}'^{cm}_2}{(2\pi)^3}\nonumber\\
&\cdot ~i{\cal T}^J_{\lambda_1,\lambda_2;\lambda'_1,\lambda'_2}({\rm p}^{'cm}_2)
G_0({\rm p}'^{cm}_2) {\cal A}^{J}_{\lambda'_1,\lambda'_2,\lambda_3;\lambda}({\rm p}'^{cm}_2,\Omega^{cm}_2).
\end{align}
After introducing partiy, the distribution can be further rewritten as~\cite{He:2017lhy,Ding:2023yuo}
\begin{align}
{d\Gamma\over dm_{12}}=&\frac{1}{(2\pi)^5}\frac{{\rm p}_1^{cm}{\rm p}_3}{16m_{B^0}}\sum_{\lambda_1,\lambda_2,\lambda_3;\lambda;J^P}|{\cal M}^{J^P}_{\lambda_1,\lambda_2,\lambda_3;\lambda}(m_{12})|^2,\label{Eq: IM}
\end{align}
with
\begin{align}
&{\cal M}^{J^P}_{\lambda_1,\lambda_2,\lambda_3;\lambda}(m_{12})\nonumber\\
&={\cal A}^{J^P}_{\lambda_1,\lambda_2,\lambda_3;\lambda}(m_{12})+\sum_{\lambda'_1,\lambda'_2}\int \frac{{\rm p}'^{cm2}_2 d{\rm p}'^{cm}_2}{(2\pi)^3}\nonumber\\
&\cdot i{\cal T}^{J^P}_{\lambda_1,\lambda_2,\lambda_3;\lambda'_1,\lambda'_2}({\rm p}'^{cm}_2,m_{12}) G_0({\rm p}'^{cm}_2) {\cal A}^{J^P}_{\lambda'_1,\lambda'_2;\lambda}({\rm p}'^{cm}_2,\Omega^{cm}_2).\label{Eq: IM0}
\end{align}
The Eq.~(\ref{Eq: IM0}) can be expressed in matrix form as $ M = A + TG A$ using the same discretization method from Eq.~(\ref{Eq: matrix}), where $T$ satisfies $T = V + VGT$. The rescattering amplitude $T$ can be solved as ${ T}=(1-{V}G)^{-1} V$. To focus on the pole of the rescattering amplitude, we need to find the position where $|1-{ V}G|=0$, with $z=E_R + i\Gamma/2$ corresponding to the total energy and width in the complex energy plane. In this study, we employ an explicit method to find the pole, as described in Ref.~\cite{Roca:2005nm}. Specifically, for each channel, we assign positive signs to the imaginary parts of the momenta $p''_o$ in the propagator below the threshold and negative signs above it. This approach yields pole positions and half-widths that are closer to those of the corresponding Breit-Wigner forms on the real axis. Consequently, for energies below the lowest threshold, we can identify possible pure bound states, while for states above the lowest threshold, the imaginary part will be obtained.

\section{Numerical Results and Discussions}\label{Sec:results}

In the three-body decay of the $B^0$ meson, potential intermediate molecular states of $N\bar{D}^{*}$ or $\bar{N}\bar{D}^{*}$ can be investigated through the invariant mass spectra of $p\bar{D}^0$ and $\bar{p}\bar{D}^0$, respectively. We begin by presenting the single-channel results for $\bar{N}\bar{D}^{*}$ and $N\bar{D}^{*}$ to provide an initial understanding of these molecular states. This approach helps us identify the existence of bound states and further determine the isospin and quantum numbers of the molecular states. We then focus on these molecular states and examine them in the $\bar{p}\bar{D}^0$ and $p\bar{D}^0$ invariant mass spectra of the decay $B^0 \rightarrow \bar{D}^0 p \bar{p}$ through a coupled-channel calculation. Additionally, we will consider possible structures near the $\bar{p}\bar{D}^0$ and $p\bar{D}^0$ thresholds to discuss the effects of potential $N\bar{D}$ or $\bar{N}\bar{D}$ states.

\subsection{Molecular states in $\bar{p}\bar{D}^0$ invariant mass spectrum}

In the invariant mass spectrum, we may observe the molecular state resulting from the $\bar{N}\bar{D}^*$ interaction, which corresponds to the experimentally observed $\Lambda_c(2940)$. Initially, we investigate whether bound states can be produced in their S-wave single-channel interactions. The explicit single-channel results of the $\bar{N}\bar{D}^*$ interactions are provided in Table~\ref{Tab:single2}.
\renewcommand\tabcolsep{0.3cm}
  \renewcommand{\arraystretch}{1.6}
  \begin{table}[h!]\footnotesize
  \begin{center}
  \caption{The binding energies of bound states from the $\bar{N}\bar{D}^*$ interaction at different $\alpha$ in a range from 2.9 to 4.1. Here, ``$--$'' means that the bound state has a binding energy larger than 50~MeV. The parameter $\alpha$ and the binding energy are in units of 1 and MeV respectively.~\label{Tab:single2}
  \label{Tab:single2}}
  \begin{tabular}{cccccccccccc}\bottomrule[1pt]
  $\alpha$    &$2.9$  &$3.1$ &$3.3$       &$3.5$      & $3.7$     & $3.9$     & $4.1$   \\\hline
 $0(3/2^{+})$&$0.6$   &$1.1$ &$5.0$       &$12.0$     &$22.6$     & $37.4$    & $--$   \\\toprule[1pt]
  \end{tabular}
  \end{center}
\end{table}

\begin{figure}[h!]
\centering
\includegraphics[scale=1.4,bb=130 140 300 235]{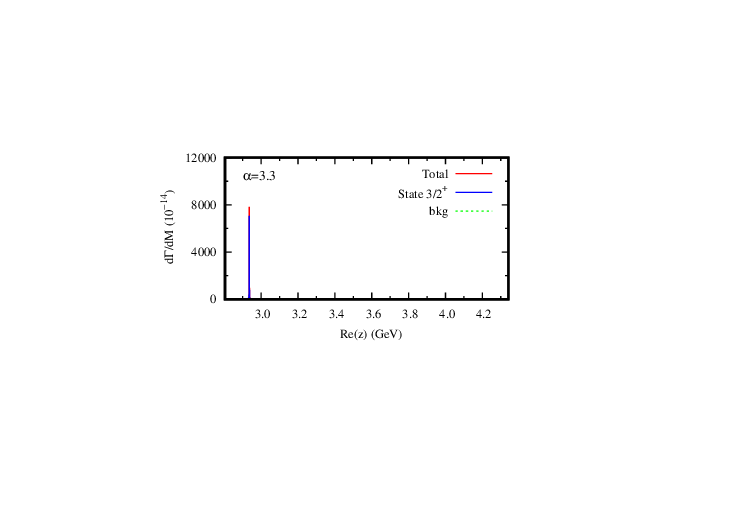}
\caption{The $\bar{p}\bar{D}^0$ invariant mass spectrum for the $B \rightarrow
\bar{D}^0 p \bar{p}$ decay process, ranging from 2.802 to 4.341~GeV with
$\alpha=3.3$. The red (solid),blue (solid), and green (dotted) curves represent
total, $3/2^+$ state, and background contributions, respectively. Note that the
red solid curve for the total result overlaps with the blue solid line for the
$3/2^+$ state, while contributions for the background (green dotted curve) is too small to
be visible.}
\label{pbd0a}
\end{figure}

As shown in Table~\ref{Tab:single2}, only the isoscalar $\bar{N}\bar{D}^*$ interaction with $J^P=3/2^+$ is found to be bound at $\alpha=2.9$, considering the exchanges of $\pi$, $\eta$, $\rho$, $\omega$, and $\sigma$ mesons. The isoscalar state with $1/2^+$ and the isovector states with $1/2^+$ and $3/2^+$ remain unbound under these conditions. Consequently, the observation of the isoscalar state with $3/2^+$ in the $\bar{p}\bar{D}^0$ mass distribution is expected after considering the coupling between the $\bar{N}\bar{D}^*$ and $\bar{N}\bar{D}$ channels. This state will subsequently decay into $\bar{p}\bar{D}^0$. In Fig.~\ref{pbd0a}, we first present the $\bar{p}\bar{D}^0$ invariant mass spectrum over the entire allowed region from 2.803 to 4.341 GeV, with a cutoff of $\alpha = 3.3$.

\begin{figure}[h!]
\centering
\includegraphics[scale=0.85,bb=90 57 368 296]{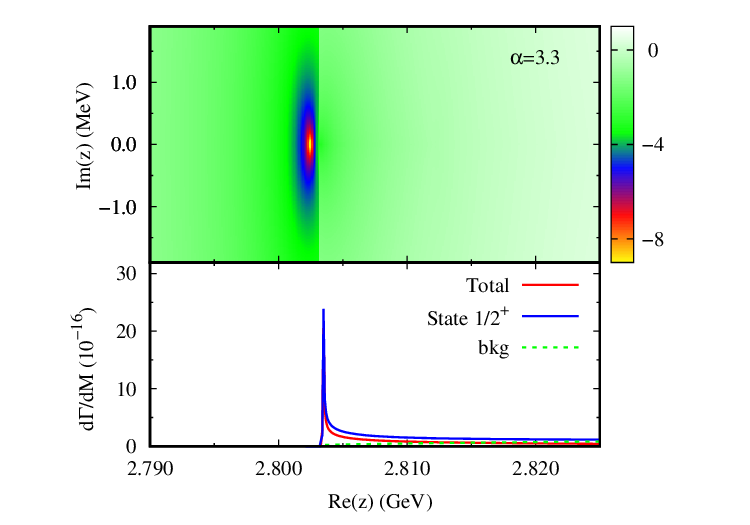}
\caption{The pole of isosacalar $\bar{N}\bar{D}$ state with $1/2^+$ (upper panel) from 2.79 to 2.825~GeV in the $\bar{p}\bar{D}^0$ invariant mass spectrum for the $B \rightarrow \bar{D}^0 p \bar{p}$ decay process with cutoff $\alpha=3.3$ (lower panel). The red (solid), blue (solid), and green (dotted) curves are for total, state, and background contributions.}
\label{pbd0lowarea}
\end{figure}

To search for other possible structures in the lower energy region, such as the
molecular states near the $\bar{N}\bar{D}$ threshold that could be candidates
for the antiparticles of $\Lambda_c(2595)$ and
$\Lambda_c(2625)$~\cite{Hofmann:2005sw,Garcia-Recio:2008rjt,Liang:2014kra,Lu:2014ina},
we provide additional invariant mass spectra within the range of 2.79 to
2.825~GeV, along with the pole of the isoscalar $\bar{N}\bar{D}$ molecule at a
cutoff $\alpha=3.3$ in Fig.~\ref{pbd0lowarea}.

As shown in the upper panel of Fig.~\ref{pbd0lowarea}, a single pole near the
lowest channel, $\bar{N}\bar{D}$, is observed on the real axis due to the
absence of open decay channels in its energy range. The mass spectrum in
Fig.~\ref{pbd0lowarea} reveals a half-peak structure with a relatively small
yield around 2.803 GeV, coinciding with the $\bar{N}\bar{D}$ threshold. This
small yield is partly due to phase space suppression, as the pole lies below the
$\bar{N}\bar{D}$ threshold, which is the observation channel considered in this
work. The existence of such a structure is insufficient to suggest that the
antiparticles of the $\Lambda_c(2595)$ and $\Lambda_c(2625)$ candidates can be
observed in the specific $B^0\rightarrow \bar{D}^0 p \bar{p}$ process for
several reasons. First, the signal of this state, with an order of magnitude of
$10^{-16}$, is insignificant in the overall process. Second, the complete
structure of this state cannot be fully captured in the invariant mass spectrum,
as the spectrum below 2.803 GeV is unphysical here. Nonetheless, the results do
not exclude the possibility that the $\bar{N}\bar{D}$ molecular state may still
have a good chance of being discovered in other processes.  Additionally, the total results are slightly
smaller than the contribution from the state with $1/2^+$, which can be
attributed to the interference between the contributions from the $1/2^+$ state
and the background.

\begin{figure*} \centering
\includegraphics[scale=1.48,bb=66 56 405 275]{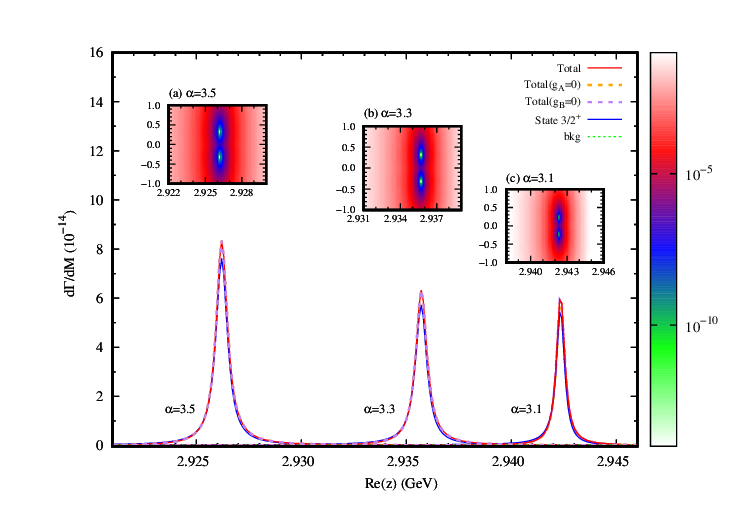}

\caption{The poles of isoscalar $\bar{N}\bar{D}^{*0}$ states with $3/2^+$ (panel
a,b,c) and the $\bar{p}\bar{D}^0$ invariant mass spectrum for $B \rightarrow
\bar{D}^0 p \bar{p}$ decay process (lower panel) with $\alpha=3.1$, 3.3 and 3.5.
The red (solid), orange (dotted), purple (dotted), blue (solid), and green
(dotted) curves are for the total, total($g_{A}=0$), total($g_{B}=0$), $3/2^+$
state, and background contributions as in Eq.~(\ref{Eq: IM0}). Note that the red
solid curve for the total result overlaps with the orange and purple dashed lines for $g_A=0$
and $g_B=0$, and the blue line for
$3/2^+$ state, while the contribution for the background (green dotted curve) is too small to
be visible.}

~\label{pbdb}
 \end{figure*}

From the entire mass spectrum in Fig.~\ref{pbd0a}, we observe a prominent structure around 2.93 GeV, which is significantly more pronounced than contributions from other regions. This distinct signal in the $\bar{p}\bar{D}^0$ invariant mass spectrum likely corresponds to the $\bar{N}\bar{D}^*$ molecular state with $I(J^P) = 0(3/2^+)$. To further analyze this peak structure, we present the poles of the state at cutoffs $\alpha = 3.1$, 3.3, and 3.5, along with the $\bar{p}\bar{D}^0$ mass spectrum from 2.941 to 2.946 GeV, as illustrated in Fig.~\ref{pbdb}. To align with the PDG value of the decay width for $B \rightarrow \bar{D}^0 p \bar{p}$, we apply an additional experimental normalization factor of $8.1 \times 10^{-4}$, considering the full mass spectrum from 2.803 to 4.341 GeV.

In Fig.~\ref{pbdb}, distinct red and blue solid peak structures appear at
approximately 2942, 2936, and 2926 MeV in the $\bar{p}\bar{D}^0$ invariant mass
spectrum, corresponding to the isoscalar bound states of $\bar{N}\bar{D}^{}$
produced at cutoffs $\alpha=3.1$, 3.3, and 3.5, respectively. The red solid
peaks predominantly arise from the contribution of the molecular state. 
Furthermore, we examine the sensitivity of our results to the numerical values
of $g_A$ and $g_B$ by considering extreme scenarios where either $g_A=0$ or $g_B=0$. This analysis is also depicted in Figure~4. It is observed that, after
scaling to the experimental decay width, both the orange and purple dotted
curves for $g_A=0$ and $g_B=0$ contributions closely overlap with the red
curve obtained with $g_A=g_B=1$ in the current study. This result
underscores that different choices of values of $g_A$ and $g_B$ does not significantly
impact the outcomes due to the normalization procedure employed.

Unlike single-channel
calculations, the single pole of the bound state deviates from the real axis
into the complex energy plane, acquiring an imaginary part and forming a pair of
conjugate poles. The small width of the poles is clearly less than 1 MeV. As the
parameter $\alpha$ increases, the peaks and poles move away from the
$\bar{N}\bar{D}^{}$ threshold. The line shapes are consistent with the narrow
peak structures of $\Lambda_c(2940)$, which can be interpreted as the 
$N{D}^{*}$ bound state with $I(J^P)=0(3/2^-)$, observed in the $p{D}^0$ mass spectrum resulting from
the $p\bar{p} \rightarrow \Lambda_c D^0 p$ process in our previous
work~\cite{He:2011jp}. When compared with experimental data of the
$\Lambda_c(2940)$, the binding energy in our calculations aligns with the
experimental results, albeit with a smaller width.

The current results reveal a clear peak for the different cutoffs considered, indicating that the direct contributions are likely to be much smaller than the rescattering contributions. If these two contributions dominate in the $B\rightarrow \bar{D}^0 p \bar{p}$ decay, the $\bar{N}\bar{D}^{*0}$ states should be easily observable.

\subsection{Molecular states in $p\bar{D}^0$ invariant mass spectrum}

The $p\bar{D}^0$ being a pure isovector state implies that only the isovector molecular states of the $N\bar{D}^*$ system can be explored in the $p\bar{D}^0$ mass spectrum. In the single-channel calculation of the isovector $N\bar{D}^*$ interaction with quantum numbers $1/2^-$ and $3/2^-$, we have taken into account the meson exchange of $\pi$, $\eta$, $\rho$, $\omega$, and $\sigma$. The explicit single-channel results of the isovector $N\bar{D}^*$ interaction with quantum numbers $1/2^-$ and $3/2^-$ are presented in Table~\ref{Tab:single}.

\renewcommand\tabcolsep{0.11cm}
  \renewcommand{\arraystretch}{1.6}
  \begin{table}[h!]
  \begin{center}
  \caption{
  Binding energies of isovector bound states resulting from the $N\bar{D}^*$
  interaction with quantum numbers $1/2^-$ and $3/2^-$ at different cutoffs
  $\alpha$ ranging from 1.1 to 3.1. Here, ``$--$'' indicates that the bound state
  has a binding energy greater than 50 MeV. The parameter $\alpha$ and the binding energy are in units of 1 and MeV respectively.
  \label{Tab:single}}
  \begin{tabular}{cccccccccccc}\bottomrule[1pt]
  $\alpha$    &$1.1$  &$1.3$ &$1.5$       &$1.7$      & $1.9$     & $2.1$     & $2.3$  &$2.5$  &$2.7$ &$2.9$  \\\hline
  $1(1/2^{-})$&$1.3$ &$3.3$  & $5.7$      &$9.0$     & $12.5$    &$16.5$   &$21.0$ &$26.3$ &$32.8$  &$41.0$ \\
  $1(3/2^{-})$&$1.3$ &$4.2$  & $8.0$      &$12.5$     & $17.7$    &$23.6$   &$30.1$ &$37.3$ &$45.1$  &$--$  \\\toprule[1pt]
  \end{tabular}
  \end{center}
\end{table}

From the results in Table~\ref{Tab:single}, it is evident that isovector bound states can be formed from the $N\bar{D}^*$ interaction with both $1/2^-$ and $3/2^-$ at a cutoff of approximately 1.1. Moreover, their binding energies gradually increase with the rise of the cutoff $\alpha$. These results indicate a significant attraction between the nucleon $N$ and the vector charmed meson $\bar{D}^*$ with $1{(1/2^-)}$ and $1{(3/2^-)}$. Additionally, the bound state with $3/2^-$ exhibits larger binding energies than the state with $1/2^-$ at a larger value of $\alpha$.

\begin{figure}[h!]
\centering
\includegraphics[scale=1.4,bb=130 140 300 235]{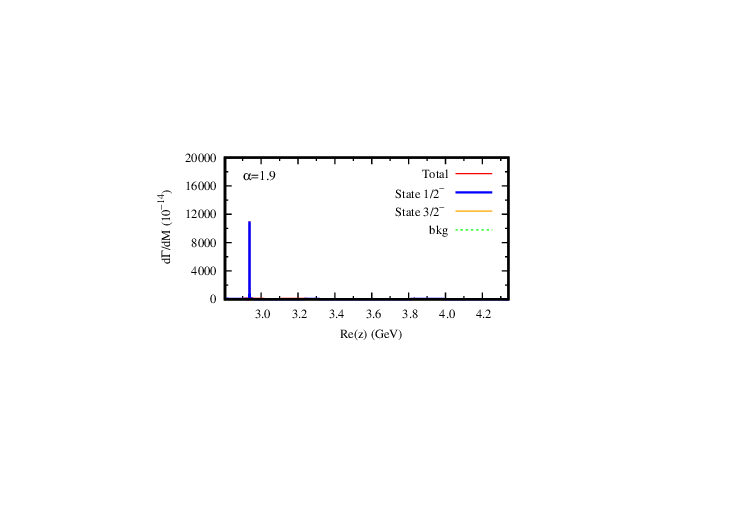}
\caption{The $p\bar{D}^0$ invariant mass spectrum for the $B \rightarrow
\bar{D}^0 p \bar{p}$ decay process from 2.803 to 2.935 GeV with $\alpha = 1.9$.
The curves in red (solid), blue (solid), orange (solid), and green (dotted)
correspond to total, $1/2^-$ state, $3/2^-$ state, and background contributions.
Note that the red solid curve for the total result overlaps with the blue solid
line for the $1/2^-$ state, while contributions for $3/2^-$ state (orange solid curve) and
the background (green dotted curves) are too small to be visible.}
\label{pdball}
\end{figure}

Both isovector $N\bar{D}^*$ bound states with quantum numbers $1/2^-$ and $3/2^-$ can be observed in the $p\bar{D}^0$ invariant mass spectrum. To begin with, we present the complete $p\bar{D}^0$ invariant mass spectrum from 2.802 to 4.341 GeV with $\alpha = 1.9$, as shown in Fig.~\ref{pdball}. In this spectrum, a structure around 2.935 GeV is observed, primarily originating from the $1/2^-$ state, which is significantly more pronounced than contributions from other regions. Similarly to the $\bar{p}\bar{D}^0$ case, we performed a detailed scan of the $p\bar{D}^0$ invariant mass spectrum in the low-energy region and found no other molecular state structures. Consequently, no specific invariant mass spectrum for the low-energy region is provided.

\begin{figure*} \centering
\includegraphics[scale=1.45,bb=66 56 405 275]{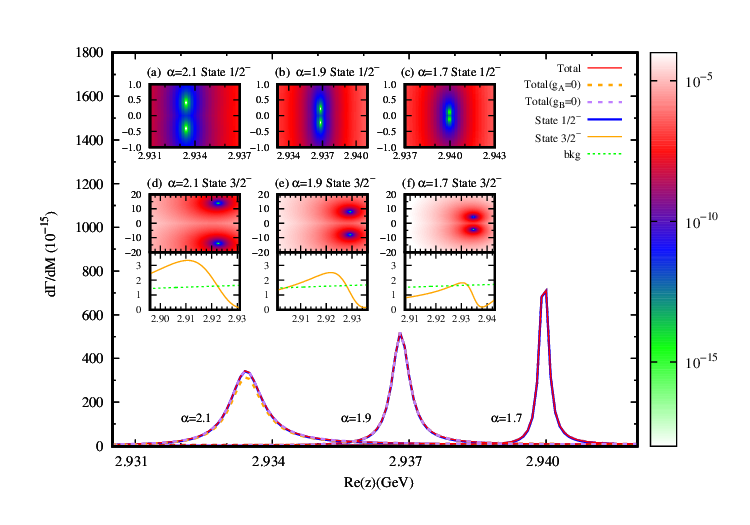}
\caption{The poles of isovector $N\bar{D}^{0}$ states with $1/2^-$ (panels a, b,
c) and $3/2^-$ (panels d, e, f) in the $p\bar{D}^0$ invariant mass spectrum for
the $B \rightarrow \bar{D}^0 p \bar{p}$ decay process (lower panel). The curves
in red (solid), blue (solid), orange (solid), and green (dotted) correspond to
the total, total($g_{A}=0$), total($g_{B}=0$), $1/2^-$ state, $3/2^-$ state, and
background contributions, respectively, and are presented for different values
of $\alpha = 1.7$, 1.9, and 2.1. Note that the red
solid curve for the total result overlaps with the orange and purple dashed lines for $g_A=0$
and $g_B=0$, and the blue line for
$1/2^-$ state, while contributions for $3/2^-$ state (orange solid curve) 
and the background (green dotted curve) are too small to
be visible.} \label{pdb0} \end{figure*}

In Fig.~\ref{pdb0}, we present the poles of isovector bound states, along with the $p\bar{D}^0$ invariant mass spectrum in the 2.93 to 2.95 GeV range, providing a more detailed analysis. Given the limited experimental data for direct comparison and the absolute distribution values derived, we present the results with varying values of $\alpha$, specifically $1.7$, $1.9$, and $2.1$, for a comprehensive discussion. An additional experimental normalization factor of $2.9 \times 10^{-5}$ should also be applied to the $p\bar{D}^0$ mass distribution.

Enhancement structures are observed around 2940, 2937, and 2933 MeV in the
$p\bar{D}^0$ invariant mass distribution. These correspond to the total
contributions and signals of the $1/2^-$ state at cutoffs $\alpha=1.7$, $1.9$,
and $2.1$, respectively. Note that both orange and purple dotted curves for $g_A
= 0$ and $g_B = 0$ overlap with the red total
contribution by after considering normalization procedure. The contributions
from the $3/2^-$ state and background are minimal, consistently remaining orders
of magnitude smaller than those from the $1/2^-$ state. These results suggest
that the signals of the isovector $1/2^-$ state can be effectively distinguished
from the background, while the $3/2^-$ state proves challenging to observe in
the $B^0 \rightarrow \bar{D}^0 p \bar{p}$ three-body decay process.

The poles of the $1/2^-$ and $3/2^-$ states are presented in the panels (a, b,
c, d, e, f) in Fig.~\ref{pdb0} for $\alpha=2.1$, $1.9$, and $1.7$, respectively.
The binding energy at $\alpha=1.7$ is marginally greater than the value obtained
from the single-channel calculation in Table~\ref{Tab:single}, and the width is
exceptionally narrow, aligning with the sharp line shape observed in the peak of
the invariant mass spectrum. As the cutoff $\alpha$ increases, the poles
gradually move towards lower energy regions, accompanied by an increase in
width. Consequently, the line shapes broaden but remain discernible in the
$p\bar{D}^0$ invariant mass distribution. The poles of the $3/2^-$ state, along
with its contributions relative to the background contributions, are presented
in panels (d, e, and f). The energies of the conjugate poles of the $3/2^-$
states are slightly smaller than the corresponding values in the single-channel
calculations from Table~\ref{Tab:single}, with a width of approximately 10 MeV.
The $3/2^-$ state decays to the final-state $p\bar{D}^0$ through a D-wave
interaction, causing its signals to be submerged by the dominant $1/2^-$
signals, making them challenging to observe.

\section{Summary}\label{Sec: summary}

The aim of this study is to explore the feasibility of observing open-charm pentaquark molecular states, specifically $N\bar{D}^*$ and $\bar{N}\bar{D}^*$, through the $B^0 \rightarrow \bar{D}^0 p \bar{p}$ three-body decay process. Utilizing a qBSE approach, we calculate the invariant mass spectra of $p\bar{D}^0$ and $\bar{p}\bar{D}^0$ to predict potential signals associated with these molecular states.

For the $\bar{N}\bar{D}^*$ system, only the isoscalar bound state with $3/2^+$ can be generated from the S-wave single-channel interaction. Notably, this state acts as the antiparticle partner of the experimentally observed $\Lambda_c(2940)$. A clear signal of this $\bar{N}\bar{D}^*$ molecular state is identified in the $\bar{p}\bar{D}^0$ mass spectrum. Although the $\bar{N}\bar{D}$ molecular state with $0(1/2^+)$ also exists, its small yield suggests that the $B^0 \rightarrow \bar{D}^0 p \bar{p}$ process considered in this work is not ideal for studying such states, which may instead be better explored in other processes.

In the $p\bar{D}^0$ invariant mass spectrum, our results reveal significant peak structures associated with the $1/2^-$ state near the $N\bar{D}^*$ threshold, while the $3/2^-$ state and background contributions remain minimal, consistently staying orders of magnitude lower than those of the $1/2^-$ state. This makes it challenging to detect signals of the $3/2^-$ state within the $p\bar{D}^0$ invariant mass spectrum.

In our calculations, the coupling constant for the direct three-body decay is determined based on the decay width of $B^0 \rightarrow \bar{D}^0 p \bar{p}$. The results highlight the importance of rescattering in the decay process, which necessitates a smaller coupling constant to adjust the differential decay width to align with experimental measurements. Additionally, in this work, we have disregarded other potential mechanisms, such as the triangle loop diagram, which could significantly contribute to the decay width and produce peaks in the invariant mass spectrum.

Nevertheless, distinct signals of the open-charm pentaquark molecular states $N\bar{D}^*$ and $\bar{N}\bar{D}^*$ are evident in the $p\bar{D}^0$ and $\bar{p}\bar{D}^0$ mass spectra resulting from $B^0 \rightarrow \bar{D}^0 p \bar{p}$. This observation suggests that with the ongoing accumulation of $B$ meson events at Belle II and LHCb, the $B^0 \rightarrow \bar{D}^0 p \bar{p}$ process offers a promising avenue for studying open-charm molecules, meriting further attention and more precise experimental measurements.

\vskip 10pt \noindent {\bf Acknowledgement} We express our gratitude to Dr. Qi Huang for valuable discussions. This project receives support from the Postgraduate Research and Practice Innovation Program of Jiangsu Province (Grant No. KYCX22-1541), the National Natural Science Foundation of China (Grants No. 12475080 and No.12405090) and the Start-up Funds of ChangZhou University (Grant No. ZMF24020043).

\end{document}